\documentclass[aps,pra,twocolumn,preprintnumbers,amsmath,amssymb,showpacs]{revtex4}

\usepackage{amsmath,amssymb,amscd}
\usepackage{psfrag}
\usepackage[dvips]{epsfig}
\usepackage{epsfig}
\usepackage{graphicx}
\usepackage{pstricks}           % e.g. colors can be used in the text
\usepackage{epstopdf}

\usepackage{enumerate}

\usepackage{theorem}
\theorembodyfont{\rmfamily}

\theoremstyle{break}

\def\QED{~\rule[-1pt]{5pt}{5pt}\par\medskip}

\def\psiGr{\psi_{\text{Gr}}}

\def\Eq{Eq.~\eqref}
\def\Eqs{Eqs.~\eqref}

\DeclareMathOperator{\diag}{diag}

\DeclareMathOperator{\real}{Re}
\DeclareMathOperator{\imag}{Im}

\newcommand{\mc}[1]{\left[\begin{matrix} #1 \end{matrix}\right]}

\newcommand{\ppfrac}[2]{\frac{\partial #1}{\partial #2}}
\newcommand{\ddfrac}[2]{\frac{d #1}{d #2}}

\def\ie{{\it i.e.}}

\begin{document}
\title{optimal laser pulse design for transferring the coherent nuclear wave packet of H$_2^+$}

\author{Jun Zhang$^{1}$ and Feng He$^{2}$\footnote{Corresponding
    author. Email: fhe@sjtu.edu.cn}}
\affiliation{$^1$ Joint Institute of UMich-SJTU and Key Laboratory of
  System Control and Information Processing (Ministry of Education) ,
  Shanghai Jiao Tong University, Shanghai, 200240, China\\
  $^2$ Key Laboratory for Laser Plasmas
  (Ministry of Education) and Department of Physics and Astronomy, Shanghai Jiao Tong University, Shanghai,
  200240, China}

\date{\today}

\begin{abstract}
  Within the Franck-Condon approximation, the single ionization of
  H$_2$ leaves H$_2^+$ in a coherent superposition of 19 nuclear
  vibrational states. We numerically design an optimal laser pulse
  train to transfer such a coherent nuclear wave packet to the ground
  vibrational state of H$_2^+$. The simulation results show that the
  population of the ground state after the transfer is more than 91\%.
  Frequency analysis of the designed optimal pulse reveals that the
  transfer principle is mainly an anti-Stokes transition, {\it i.e.}
  the H$_2^+$ in $1s\sigma_g$ with excited nuclear vibrational states
  is first pumped to $2p\sigma_g$ state by the pulse at an appropriate
  time, and then dumped back to $1s\sigma_g$ with lower excited or ground
  vibrational states.
\end{abstract}


\maketitle

\section{Introduction}
Controlling coherent quantum states has been a longstanding goal since
the invention of laser pulses. With the rapid advent of technology in
recent years~\cite{Krausz09}, researchers can now fine-tune the laser
parameters to control the ultrafast processes inside atoms and
molecules~\cite{Lein07,Becker12,Plaja12}. For example, by varying the
relative phase of a two-color ($\omega$-$3\omega$) laser field with
$\omega$ the fundamental angular frequency, the target molecule may be
constructively or destructively excited by simultaneously absorbing
the $\omega$ and $3\omega$ photons~\cite{Brumer11}.
Another example is to change the time delay between two laser pulses
so that the molecule can be first pumped to an intermediate state,
then evolves and accumulates the time-dependent phases, and later be
dumped to a different final state, thereby changing the production of
a chemical reaction~\cite{Tannor85}. Furthermore, specific tailoring
of the laser field may dictate a complex chemical reaction to follow
one particular channel and stay away from all the others, achieving a
selective terminal state~\cite{Shi88}. Most recently, thanks to the
phase-stabilized few-cycle laser pulse \cite{Baltuska03}, the emission
of an ionized electron \cite{Paulus01}, or the charge-direct transfer
between nuclei \cite{Roudnev04,Kling06} also become possible.  The
attosecond pulse may be used to selectively excite or ionize
the target at unprecedentedly precise timing during the chemical
reactions~\cite{He07,Uiberacker07,Sansone10}, and thus helps
understanding the time-resolved fundamental physics.

\begin{figure}
\includegraphics[width=0.9\columnwidth]{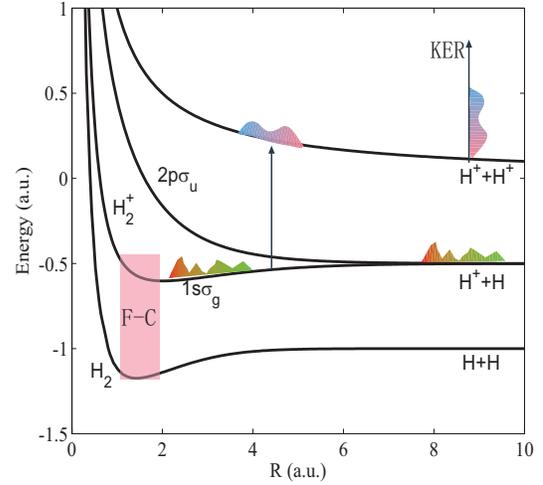}
\caption{(Color online) The schematic picture for the interaction
  between laser pulses and H$_2$. Four curves from bottom up are the
  potential curves for H$_2$, H$_2^+$ in $1s\sigma_g$, H$_2^+$ in
  $2p\sigma_u$, and coulomb explosion. F-C remarks the Franck-Condon
  transition area.  The single ionization of H$_2$ produces the H$_2^+$ in
  $1s\sigma_g$, followed by the dissociative ionization by the probe
  pulse. The kinetic energy release (KER) reflects the information
  when the coulomb explosion happens.} \label{fig1}
\end{figure}

As the simplest neutral molecule, H$_2$ (or D$_2$) is often chosen
as a prototype system to be controlled and analyzed. In the past few
decades, the basic processes for H$_2$ exposed in strong laser fields
have been extensively studied. As shown in Fig.  \ref{fig1}, after one
electron absorbs enough photon energy and escapes from the nuclei, it
is left with a molecular ion H$_2^+$ in $1s\sigma_g$
state~\cite{Ergler06}. From the Franck-Condon
approximation~\cite{Dunn68}, we can assume that initially the nuclear
wave packet (NWP) of H$_2^+$ is the same as the ground state of H$_2$,
and then it evolves along the $1s\sigma_g$ potential curve, as shown
in Fig.~\ref{fig2}(c)~\cite{Thumm03}. If a time-delayed probe pulse is
subsequently introduced, H$_2^+$ may dissociate through the
laser-induced coupling between $1s\sigma_g$ and $2p\sigma_u$
\cite{Kling06,Bandrauk81,Bucksbaum90,Suzor90,Frasinski99}. The mixture
of the dissociative channels, {\it i.e.} the paired and unpaired states,
will induce an asymmetric electron
localization~\cite{Kremer09,Fischer10,Singh10,He07,He081,He12,He082,He083,Liu12,Lan12,Anis12}.
Alternatively, H$_2^+$ may also be ionized by the probe pulse, which
leads to the Coulomb explosion~\cite{Zuo95,Ergler06}. The internuclear
distance when the ionization of H$_2^+$ taking place can be reflected
by the kinetic energy release (KER) of the Coulomb-explosion
fragments~\cite{Pavicic05,Bocharova08,Manschwetus09,Staudt07}.  In
addition to these non-electron correlation processes, the first
ionized electron may come back and rescatter with H$_2^+$,
accompanying with the excitation of H$_2^+$~\cite{Niikura03,Tong03}
and high harmonic generation~\cite{Chirila08}, or
auto-ionization~\cite{Saugout07}. If the laser environments are
appropriate, the single ionization of H$_2$ may leave the H$_2^+$ in
higher excited electronic states, {\it e.g.}
$2p\sigma_u$~\cite{Sansone10} or $2p\pi_u$~\cite{Kelkensberg11}.

In these processes, the complexity of the H$_2^+$ NWP makes the whole
process even more complicated.  After the single ionization, the NWP
of H$_2^+$ is a superposition of 19 vibrational states with negligible
auto-dissociative states. Each vibrational state has a different
spatial distribution. Once the molecule is dissociated or ionized,
each vibrational state also contributes a different KER to the
molecular fragments~\cite{McKenna09}. The coherent superposition of
the vibrational states may partly smear the asymmetric electron
localization~\cite{Anis12}, or lead to the time-dependent dissociation
\cite{Niikura06}. Since the stationary nuclear state can significantly
simplify the physical picture, it is often desired to transfer the
coherent NWP to one stationary state, especially to the ground
vibrational state of H$_2^+$.

For the hetero-nuclear molecular ion HD$^+$, the permanent dipole
induced by the asymmetric nuclear mass may transfer the Franck-Condon
NWP to the ground vibrational state~\cite{Orr07}. However, for
H$_2^+$, the external laser field has to be applied. Niederhausen and
Thumm suggested to use the multi-pulse protocol to control the
Franck-Condon coherent NWP and found that in the final coherent
population, the largest proportion for a certain vibrational state can
exceed 60\%~\cite{Niederhausen08}. Niikura {\it et al.} studied to
exert a laser-induced dipole force at an appropriate time to achieve
up to 50\% population for the ground vibrational
state~\cite{Niikura04}. Picon {\it et al.} proposed to use a chirped
few-hundred-femtosecond pulse or pulse train to transfer the first and
second excited vibrational states to the ground vibrational state with
the proportion up to 90\%~\cite{Picon11}. Bryan {\it et al.} used the
pump-modify-probe strategy to manipulate the vibrational states, where
the time-delayed second pulse may modify the relative populations of
different states~\cite{Byran11}.

In this paper we use optimal control theory to numerically design a
laser pulse train to tailor the coherent vibrational states. We
formulate it as a minimax problem with bounded constraints, and then
apply sequential linear programming algorithm~\cite{Zhang:12} to solve
it. The gradient of the performance metric with respect to the laser
pulses can be derived in an analytic manner, which facilitates the
numerical computation. For the initial Franck-Condon NWP, the transfer
to the ground vibrational state is achieved with the population more
than 91\%.

\section{numerical model}
\subsection{Two-state equation}
\begin{figure}
\includegraphics[width=1\columnwidth]{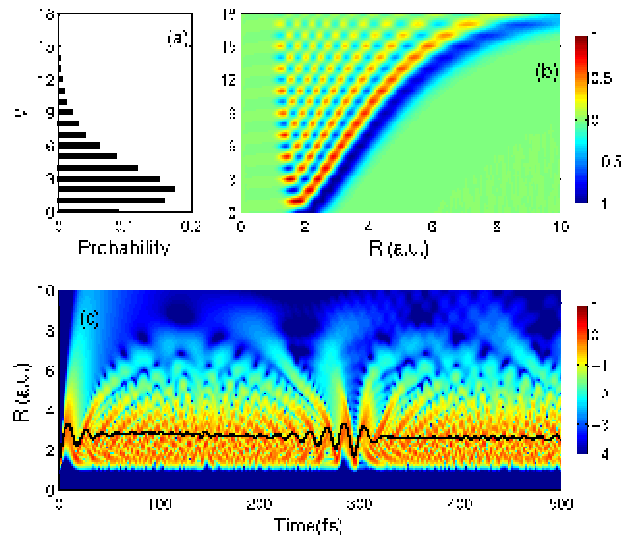}
\caption{(Color online) (a) Franck-Condon coefficients of the
  NWP of H$_2^+$. (b) Wave functions for the 19 vibrational states (in
  linear scale). (c) The free propagation of the Franck-Condon NWP of
  H$_2^+$ in the $1s\sigma_g$ potential curve (in logarithmic scale).}
\label{fig2}
\end{figure}

Consider the case that the single ionization of H$_2$ produces a free
electron and a molecular ion H$_2^+$ in $1s\sigma_g$, where the NWP of
H$_2^+$ is described by the Franck-Condon approximation. If the
time-delayed probe pulse is introduced to cause the dissociation, the
dynamics is mainly governed by a two-state equation (atomic units are
used unless otherwise stated)
\begin{equation}
  \label{eq:3}
  i\ppfrac{}{t}\mc{\psi_g(R,t)\\ \psi_u(R,t)}=
\mc{T_R+V_g(R)& d_{gu}(R)E(t)\\
d_{gu}(R)E(t)& T_R+V_u(R)} \mc{\psi_g(R,t)\\ \psi_u(R,t)},
\end{equation}
where $\psi_g(R,t)$, $\psi_u(R,t)$ are the NWP corresponding to the
electron in $1s\sigma_g$ and $2p\sigma_u$ states, and $V_g(R)$,
$V_u(R)$ are the potential curves for $1s\sigma_g$ and $2p\sigma_u$
states, respectively. The dipole coupling between these two states is
represented by $d_{gu}$, and $T_R=-\frac{1}{2M}\ppfrac{^2}{R^2}$ is
the second order differential operator, where $M=918$ is the reduced
mass of two nuclei. The molecular rotation is neglected since we limit
the pulse duration within a few tens of femtoseconds.  The initial NWP
is given by
\begin{equation}
  \label{eq:5}
  \psi_g(R,0)=\psiGr^0(R), \quad \psi_u(R,0)=0,
\end{equation}
where $\psiGr^0(R)$ is the ground state of H$_2$. We use the
Split-Operator method to solve \Eq{eq:3}~\cite{Feit82}. The $R$ spans
from 0 to 40, and the spatial step $\Delta R=0.04$. The time step is
set as $\Delta t=1$. Mask functions are used to suppress the
unphysical reflection by the boundary of the simulation box.

Our objective is to design a laser pulse $E$ such that at the terminal
time $T_f$, $\psi_g$ can be transferred to the ground state of
H$_2^+$, \ie
\begin{equation}
  \label{eq:8}
  \psi_g(R,T_f)=\psi_g^{\nu=0}(R),
\end{equation}
where $\nu$ is the index of the vibrational state.  The initial NWP
$\psi_g(R,0)$ is mainly a superposition of $19$ vibrational states. By
projecting it to the vibrational eigenstates of H$_2^+$, we obtain the
Franck-Condon coefficients, as shown in Fig.~\ref{fig2}(a).  The three
vibrational states $\nu=1$, $2$, and $3$ amount to around 50\% of the
total population.  Fig.~\ref{fig2}(b) shows the wave function for all
these 19 vibrational states, and Fig.~\ref{fig2}(c) plots
$|\psi_g(R,t)|^2$. The black curve in Fig.~\ref{fig2}(c) is the
expected time-dependent internuclear distance $\langle R(t)\rangle$.
Clearly, the NWP goes through a collapse and revival procedure, and
the revival time is about 300 fs \cite{ Ergler06,Thumm03}.

\subsection{Numerical optimization algorithm}
We formulate the design of a laser field $E$ to realize the NWP
transfer of $\psi_g$ as a minimax problem, and then apply a sequential
linear programming algorithm to solve it. To avoid ionization of
H$_2^+$, we restrict the amplitude of $E$ within $0.1$, and the pulse
duration less than 32 fs.

The wave function transfer is formulated as a constrained minimax
problem on the laser electric field $E$:
\begin{equation}
  \label{eq:4}
\min_{E} \max_{n\in\{0, 1, \cdots, N-1\}} J_n,
\end{equation}
subject to
\begin{equation}
  \label{eq:20}
|E(t_k)| \le 0.1,\ k=0, \cdots, K-1,
\end{equation}
where
\begin{equation}
\label{eq:23}
J_n=\frac12 \left\|\psi_g^0(R_n) -e^{-i\alpha} \psi_g(R_n, T_f)\right\|^2.
\end{equation}
Here $n$ (or $k$) is the index for the spatial (or temporal) step, $N$
(or $K$) is the total points in the spatial (or temporal) axis, and
$\alpha$ is a global phase to be determined soon.  The function $J_n$
quantifies the difference between the desired and actually achieved
wave functions at the spatial grid $R_n$. If the maximum error of
$J_n$ is minimized over the whole spatial range, one can expect that
the desired wave function is achieved.

The NWP transfer fidelity can be measured by
\begin{equation}
  \label{eq:10}
F=\real \left\{e^{i\alpha}\left\langle \psi_g(R,T_f) |
\psi_g^{\nu=0}(R) \right\rangle \right\}.
\end{equation}
The global phase $\alpha$ in \Eqs{eq:23} and~\eqref{eq:10} can be
obtained by maximizing the fidelity:
\begin{equation}
  \label{eq:12}
\alpha=-\arg \left\{\left\langle \psi_g(R,T_f) |
\psi_g^{\nu=0}(R) \right\rangle \right\},
\end{equation}
where $\arg$ denotes the argument of a complex number.

The minimax problem has been extensively studied in the optimization
and control community~\cite{Bandler:79, Hald:81, Conn:92}.  To find
the optimal laser field, we start from an initial guess and then
gradually approach the optimal solution by iteration. Suppose that at
the $j$-th iteration, the current laser pulse is $E^j$. We need to
determine a small increment $\Delta E^j$ such that at the $(j+1)$-th
step, the new laser pulse $E^{j+1}=E^j+\Delta E^j$ is a better
solution to minimize the transfer error $J_n$. By first order
approximation, we have
\begin{equation}
 J_n(E^{j+1})\approx J_n(E^j)+\nabla_{E^j}^T J_n(E^j) \Delta E^j.
\end{equation}
The analytic derivation of the gradient $\nabla_{E^j}^T J_n(E^j)$ is
given in the appendix.

We then apply a sequential linear programming algorithm as follows:
{\it \begin{enumerate}
\item Choose a small constant as the initial guess of the electric
  field;
\item  At the $j$-th step, compute $ J_n(E^j)$ and $\nabla_{E^j} J_n(E^j)$;
\item Determine the increment $\Delta E^j$ from the following linear
  programming problem:
  \begin{equation*}
    \min_{\Delta E^j} \gamma,
  \end{equation*}
subject to
\begin{eqnarray*}
&&\nabla_{E^j}^T J_0(E^j) {\Delta E^j} + J_0(E^j)\le \gamma,\nonumber \\
 &&\qquad \quad \vdots\nonumber\\
&&\nabla_{E^j}^T J_{N-1}(E^j) {\Delta E^j}+ J_{N-1}(E^j) \le \gamma,\nonumber \\
&&-0.1 -E^j\le \Delta E^j\le 0.1-E^j.\nonumber\\
\end{eqnarray*}
\item  Let $E^{j+1}=E^j+\epsilon \Delta E^j$, where $\epsilon$ is a
  small positive number controlling the step size;
\item  Repeat Steps (2)--(4) until a desired convergence is reached.
\end{enumerate}}
Note that in each iteration we only need to solve a linear programming
problem, which can be readily calculated by numerical packages.

\section{Optimal design results}
\begin{figure}[tb]
\includegraphics[width=0.9\columnwidth]{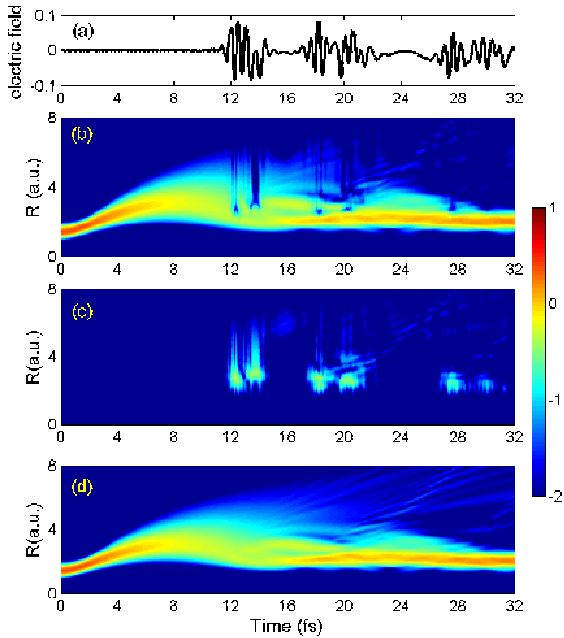}
\caption{(Color online) (a) The designed
  optimal laser pulse train. The evolution of $|\psi_g(R,t)|^2$ (b),
  $|\psi_u(R,t)|^2$(c), and $|\psi_g(R,t)|^2+|\psi_u(R,t)|^2$ (d) (all
  in logarithmic scale).  }
\label{fig3}
\end{figure}

Starting from $E(t)=0.01$, after around 12,000 iterations and 350+
hours computation on a desktop computer with Intel i5 CPU, we have
obtained a satisfactory optimal laser pulse as shown in
Fig.~\ref{fig3}(a).  Fig.~\ref{fig3}(b), (c), (d) show the optimal
laser pulse induced NWP evolution for $|\psi_g(R,t)|^2$,
$|\psi_u(R,t)|^2$, and $|\psi_g(R,t)|^2+|\psi_u(R,t)|^2$,
respectively.

Fig. \ref{fig3}(a) reveals an interesting physical story. First of
all, the optimal laser field is a pulse train. After the inception of
H$_2^+$, nearly no electric field is introduced until $t=11$ fs. In
this period, the NWP propagates freely to the outer turning point and
then turns back, as shown in Fig.~\ref{fig3}(b). The main pulse
appears at around $t=11$ fs, at which time the NWP is moving inward
instead of outward. This is important because if the laser field
starts interacting with the NWP when it is moving outward, part of the
wave packets will directly dissociate and the subsequent laser pulse
has little chance to pull them back to the bound states
\cite{Niikura04}.

To transfer $\psi_g(R,t)$ to $\psi_g^{\nu=0}(R)$, $\psi_u(R,t)$ must
be mediated. From a closer look at Fig.~\ref{fig3}(b) and (c), one may
find that within each oscillation of the electric field, part of
$\psi_g$ and $\psi_u$ are exchanged. The wave
function $\psi_u(R,t)$ mainly distributes close to the range $R=3$.
The quantity $|\psi_g(R,t)|^2+|\psi_u(R,t)|^2$ gives a smooth
evolution of the NWP, as depicted in Fig.~\ref{fig3}(d). At the
terminal time, the NWP has been transferred to the ground vibrational
state.

\begin{figure}[tb]
\includegraphics[width=0.9\columnwidth]{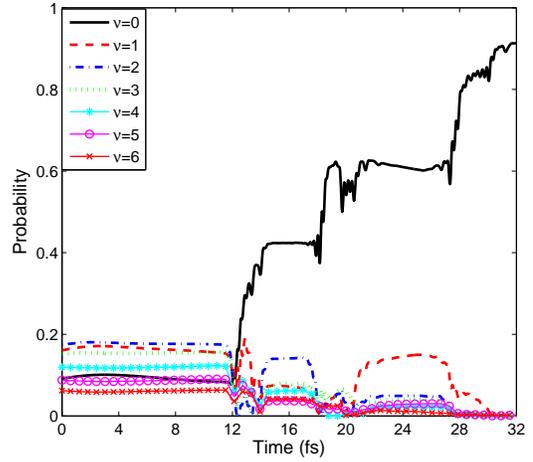}
\caption{(Color online) The time-dependent probabilities for the
  lowest 7 vibrational states. }
\label{fig4}
\end{figure}

To gain a deeper understanding of the transfer principle, we trace the
time-dependent probability evolution of each individual vibrational
state, which can be written as
\begin{equation}
\label{p_vib}
P_{\nu}(t)=\left|\langle \psi_g^{\nu}(R)| \psi_g(R,t)\rangle
\right|^2, \quad \text{for } \nu=0, \cdots, 18.
\end{equation}
Fig.~\ref{fig4} shows $P_{\nu}(t)$ for the first seven vibrational
states.  It is clear that the population of $\nu=0$ increases to 91\%
at the end of the evolution, and the staircase jumps take place at the
times when the laser pulse is introduced. The quick increasing of the
ground state population and the precipitous dropping of the excited
vibrational states indicate that the laser induced coupling is roughly
an anti-Stokes transition: H$_2^+$ with higher nuclear vibrational
states is excited from $1s\sigma_g$ to $2p\sigma_u$, and then
de-excited to $1s\sigma_g$ with lower nuclear vibrational states.
The probability evolution details show more physical
scenarios.  At the beginning $\nu=1$ and $\nu=2$ states
have similar probabilities. After the first laser pulse, the
probability of $\nu=1$ is halved, whereas the probability of $\nu=2$
does not change much. Surprisingly, after the second pulse, the
probability of $\nu=1$ is doubled and is much larger than that of
$\nu=2$.  After these two pulses, the vibrational states with $\nu\geq
2$ are already very small, and the upcoming third pulse mainly
transfers $\nu=1$ to $\nu=0$.  During the whole process, $\nu=1$ state
works as a temporary reservoir for storing some population, for
ultimately maximizing the population of $\nu=0$.

\begin{figure}
\includegraphics[width=0.9\columnwidth]{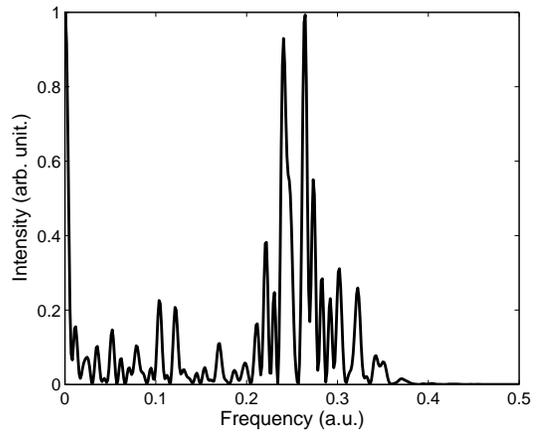}
\caption{The frequency spectrum for the designed optimal laser pulse
  train. }
\label{fig5}
\end{figure}

The frequency spectrum of the obtained laser pulse train is shown in
Fig.~\ref{fig5} after performing the Fourier transform. The main frequency
component is around 0.25. This is consistent with the optimization
result in Fig.~\ref{fig3}(c), where $\psi_u(R)$ is mainly excited at
the internuclear distance between 2.5 to 3, for the energy
gap between $1s\sigma_g$ and $2p\sigma_u$ at the corresponding internuclear distance is around 0.25. The
frequency analysis demonstrates that the multi-frequency laser pulse
train pumps and dumps H$_2^+$ with different frequency components.

\section{Conclusions}
In conclusion, by restricting the laser pulse duration to less than 32 fs
and confining the electric amplitude within 0.1, we numerically design
an optimal laser pulse train to successfully transfer the initial
Franck-Condon NWP to the ground vibrational state of H$_2^+$ with a population
91\%, and the dissociation probability is only 9\%. The optimal laser
pulse train does not act on the NWP until the NWP is moving inward.
The field-induced Raman transition between $1s\sigma_g$ and
$2p\sigma_u$ transfers the highly excited vibrational states to
$\nu=0$ directly, or indirectly first to $\nu=1$ but finally to
$\nu=0$ state. This control algorithm can be extended to other
molecules.

\begin{acknowledgments}
  Both authors thank the financial support from Shanghai Pujiang
  scholar funding (Grant No. 11PJ1405800, 11PJ1404800), NSFC (Grant
  No.  61174086, 11104180, 11175120), and Project-sponsored by SRF for
  ROCS SEM. JZ thanks the Innovation Program of Shanghai Municipal
  Education Commission (Grant No. 11ZZ20), and State Key Lab of
  Advanced Optical Communication Systems and Networks, SJTU, China.
  FH thanks the NSF of Shanghai (Grant No. 11ZR1417100) and the Fok
  Ying-Tong Education Foundation for Young Teachers in the Higher
  Education Institutions of China (Grant No.  131010).
\end{acknowledgments}

\appendix*
\section{Derivation of $\nabla_{E^j} J_n$}
For completeness, we first briefly describe the numerical
procedure to solve the Schr\"odinger equation~\eqref{eq:3}. We follow the standard split-operator techniques in
Refs.~\cite{Feit82,Tannor:07,Schwendner:97}. Let
\begin{equation*}
 \label{eq:24}
  E=[E_0, E_1, \cdots, E_{K-1}],\quad  R=[R_0, R_1, \cdots, R_{N-1}].
\end{equation*}
The solution of \Eq{eq:3} can be written as
\begin{equation}
 \label{eq:13}
 \psi(R,T_f)= \prod\nolimits_{k=0}^{K-1} e^{-i H_k \Delta t} \psi(R,0),
\end{equation}
where
\begin{equation*}
 \label{eq:14}
 H_k= \mc{T_R+V_g(R)& d_{gu}(R)E_k\\
 d_{gu}(R)E_k& T_R+V_u(R)},
\end{equation*}
which is decomposed as
\begin{equation*}
 \label{eq:15}
H_k=T+G_k,
\end{equation*}
where
\begin{equation*}
 \label{eq:16}
T=\mc{T_R& 0\\ 0& T_R}, \quad
G_k=\mc{V_g(R)& d_{gu}(R)E_k\\ d_{gu}(R)E_k& V_u(R)}.
\end{equation*}
The propagation operator $e^{-i H_k \Delta t}$ in \Eq{eq:13} can be
calculated by the split-operator method:
\begin{equation}
 \label{eq:17}
e^{-i H_k \Delta t}=e^{-i T \frac{\Delta t}2}
e^{-i G_k \Delta t}e^{-i T \frac{\Delta t}2}+O(\Delta t^3).
\end{equation}
Substitution of \Eq{eq:17} into \Eq{eq:13} yields
\begin{equation}
 \label{eq:18}
\psi(R,T_f)
= e^{-i T \frac{\Delta t}2} \left(\prod_{k=0}^{K-1}
e^{-i G_k \Delta t}e^{-i T \Delta t}\right)
e^{i T \frac{\Delta t}2} \psi(R,0).
\end{equation}
Here the terms $e^{-i T \frac{\Delta t}2}$ and $e^{-i T \Delta t}$ can
be calculated by Fast Fourier Transform (FFT)~\cite{Tannor:07}. Since
all the four blocks in $G_k$ are diagonal matrices,
we can transform $G_k$ into a block diagonal matrix
$\tilde{G}_k=\diag\{G_k^0, G_k^1, \cdots, G_k^{N-1}\}$,
where
\begin{equation}
 \label{eq:29}
 G_k^n=\mc{V_g(R_n)&d_{gu}(R_n)E_k\\ d_{gu}(R_n)E_k&V_u(R_n)}.
\end{equation}
Because $G_k^n$ is symmetric, it can be derived that
\begin{equation}
 \label{eq:6}
 \begin{aligned}
&e^{-i G_k^n \Delta t}
= \exp\left\{-i\dfrac{V_g(R_n)+V_u(R_n)}2 \Delta t\right\}
\left\{\cos \frac{\theta_k\Delta t}{2}I \right. \\
&\left.- i \sin \frac{\theta_k\Delta t}{2}
\left(\frac{V_g(R_n)-V_u(R_n)}{\theta_k} \sigma_z
+ \frac{2d_{gu}(R_n)E_k}{\theta_k}
 \sigma_x\right)\right\},
 \end{aligned}
\end{equation}
where  $\sigma_x$, $\sigma_z$ are Pauli matrices, and
\begin{equation}
 \label{eq:7}
 {\theta_k}=\sqrt{(V_g(R_n)-V_u(R_n))^2+(2d_{gu}(R_n)E_k)^2}.
\end{equation}
This completes the numerical solution of \Eq{eq:3}.

Now from~\Eq{eq:23}, we have
\begin{equation}
 \label{eq:34}
 \begin{aligned}
\nabla_{E} J_n=& -\real \bigg\{\left( \overline{\psi_g^0}(R_n)
 -e^{i\alpha} \overline{\psi_g}(R_n, T_f)\right)\\
& \qquad \qquad\times \nabla_{E}\left( e^{-i\alpha} \psi_g(R_n,
    T_f)\right) \bigg\}.
 \end{aligned}
\end{equation}
For an element $E_k$ in the vector $E$, it is easy to get
\begin{equation}
  \label{eq:1}
   \begin{aligned}
& \ppfrac{}{E_k} e^{-i\alpha} \psi_g(R_n,  T_f)\\
=&e^{-i\alpha}\ppfrac{}{E_k} \psi_g(R_n,  T_f)
-e^{-i\alpha} \psi_g(R_n,  T_f)\ppfrac{\alpha}{E_k}.
  \end{aligned}
\end{equation}
From \Eqs{eq:18}, we obtain that
\begin{equation*}
  \label{eq:2}
  \begin{aligned}
& \ppfrac{}{E_k}\psi(R,T_f)
= e^{-i T \frac{\Delta t}2}
\bigg(\prod_{l=k+1}^{K-1}  e^{-i G_{l} \Delta t}e^{-i T \Delta t}\bigg) \\
\times& \ppfrac{}{E_k}e^{-i G_k \Delta t}e^{-i T \Delta t}
\bigg(\prod_{l=0}^{k-1}  e^{-i G_{l} \Delta t}
 e^{-i T \Delta t}\bigg)
 e^{i T \frac{\Delta t}2} \psi(R,0).
  \end{aligned}
\end{equation*}
From \Eqs{eq:6} and~\eqref{eq:7}, it follows that
\begin{equation*}
  \label{eq:9}
\ppfrac{}{E_k}e^{-i G_k^n \Delta t}
=\ppfrac{}{\theta_k}e^{-i G_k^n \Delta
  t}\ppfrac{\theta_k}{E_k}
+\ppfrac{}{E_k}e^{-i G_k^n \Delta t}.
\end{equation*}
Proceeding further, we obtain
\begin{equation*}
  \label{eq:11}
    \begin{aligned}
&  \ppfrac{}{\theta_k}e^{-i G_k^n \Delta t}
= e^{-i \frac{V_g+V_u}2 \Delta t}
\bigg\{-\frac{\Delta t}2\sin \frac{\theta_k\Delta t}{2}I \\
&\qquad\qquad -\frac{i\Delta t}2 \cos\frac{\theta_k\Delta t}{2}
\left(\frac{V_g-V_u}{\theta_k} \sigma_z + \frac{2d_{gu}E_k}{\theta_k}
  \sigma_x\right) \\
&\qquad\qquad +i \sin \frac{\theta_k\Delta t}{2}
\left(\frac{V_g-V_u}{\theta_k^2} \sigma_z + \frac{2d_{gu}E_k}{\theta_k^2}
  \sigma_x\right)\bigg\},
  \end{aligned}
\end{equation*}
and
\begin{eqnarray*}
  \label{eq:36}
&&\ppfrac{\theta_k}{E_k}=\frac{4d_{gu}^2 E_k}{\theta_k},\hspace{4cm} \\
\label{eq:19}
&&\ppfrac{}{E_k}e^{-i G_k^n \Delta t}
=-i e^{-i \frac{V_g+V_u}2 \Delta t}
 \sin \frac{\theta_k\Delta t}{2}
\frac{2d_{gu}}{\theta_k} \sigma_x.
\end{eqnarray*}

Lastly, we need to calculate $\ppfrac{\alpha}{E_k}$ in \Eq{eq:1}. Rereading
\Eq{eq:12} and defining
\begin{equation*}
  \begin{aligned}
   p&=\real\left\{\sum\nolimits_{n=0}^{N-1}\psi_g^0(R_n)
   \overline{\psi_g(R_n, T_f)}\right\}, \\
q&=\imag\left\{\sum\nolimits_{n=0}^{N-1}\psi_g^0(R_n)
   \overline{\psi_g(R_n, T_f)}\right\},
  \end{aligned}
\end{equation*}
we obtain
\begin{equation*}
  \label{eq:33}
 \ppfrac{\alpha}{E_k}=\frac{1}{p^2+q^2}\left(p\ddfrac{q}{E_k}
-q\ddfrac{p}{E_k} \right),
\end{equation*}
where $\ddfrac{p}{E_k}$ and $\ddfrac{q}{E_k}$ are none other than the
real and imaginary parts of the quantity
\begin{equation*}
  \label{eq:35}
\sum\nolimits_{n=0}^{N-1}\psi_g^0(R_n) \ppfrac{}{E_k} \overline{\psi_g(R_n,
  T_f)}.
\end{equation*}

Combining all these equations, we can calculate $\nabla_{E^j}
J_n$ in an explicit manner.

\end{document}